\def\simless{\mathbin{\lower 3pt\hbox
             {$\rlap{\raise 5pt\hbox{$\char'074$}}\mathchar"7218$}}}    

\def\simmore{\mathbin{\lower 3pt\hbox
             {$\rlap{\raise 5pt\hbox{$\char'076$}}\mathchar"7218$}}}    

\documentclass[structabstract]{aa}

\usepackage{graphicx}
\begin{document}
\title{
Formation and destruction of jets in X-ray binaries
}

\subtitle{}

\author{
N. D. Kylafis\inst{1,2}, I. Contopoulos\inst{3}, D. Kazanas\inst{4},
\and
D. M. Christodoulou\inst{5}
}

\institute{
University of Crete, Physics Department \& Institute of
Theoretical \& Computational Physics, 71003 Heraklion, Crete, Greece\\
\and
Foundation for Research and Technology-Hellas, 71110 Heraklion, Crete, Greece\\
\and
Research Center for Astronomy, Academy of Athens, Athens 11527, Greece\\
\and
NASA/GSFC, Code 663, Greenbelt, MD 20771, USA\\
\and
Department of Mathematical Sciences, University of Massachusetts Lowell,
Lowell, MA 01854, USA
}

\date {Received ; Accepted ;}


\abstract
{
Neutron-star and black-hole X-ray binaries (XRBs) exhibit radio jets,
whose properties depend on the X-ray spectral state and history of
the source.  In particular, black-hole XRBs emit compact,
steady radio jets when they are in the so-called hard state. These
jets become eruptive as the sources move toward the soft state,
disappear in the soft state, and then re-appear when the sources
return to the hard state. The jets from
neutron-star X-ray binaries are typically weaker radio emitters
than the black-hole ones at the same X-ray luminosity
and in some cases radio emission is detected in the soft state.
}
{
Significant phenomenology has been developed to describe the spectral
states of neutron-star and black-hole XRBs, and there is general
agreement about the type of the accretion disk around the compact object
in the various spectral states.  We investigate whether
the phenomenology describing the X-ray emission on one hand
and the jet appearance and disappearance on the other
can be put together in a consistent physical picture.
}
{
We consider the so-called Poynting-Robertson cosmic battery (PRCB),
which has been shown to explain in a
natural way the formation of magnetic fields in the disks of AGNs
and the ejection of jets.  We investigate whether the
PRCB can also explain the formation, destruction, and
variability of jets in XRBs.
}
{
We find excellent agreement between the conditions under which the
PRCB is efficient (i.e., the type of the accretion
disk) and the emission or destruction of the radio jet.
}
{
The disk-jet connection in XRBs can be explained in a natural way
using the PRCB.
}


\authorrunning{Kylafis et al. 2011}
\titlerunning{Jets in X-ray binaries}

\maketitle


\section{Introduction}

Jets have been observed from both neutron-star and black-hole
X-ray binaries (XRBs). There have been many studies of the theory 
of jets and outflows off disks of compact objects
(e.g., Blandford \& Payne 1982; Meier 2001, 2005; Blandford \&
Begelman 2004; Machida, Nakamura, \& Matsumoto 2006; Fereira
et al. 2006; Fragile \& Meier 2009; Spruit 2010, and many others).
A vast amount of phenomenology has also been developed regarding
accretion and jets in black-hole XRBs as summarized in Sect. 3 below
(for reviews see Fender, Belloni, \& Gallo 2004; Homan \& Belloni
2005; Remillard \& McClintock 2006; Done, Gierli\'nski, \& Kubota
2007; Fender, Homan, \& Belloni 2009; Belloni 2010; Markoff 2010).
Neutron-star jets are typically fainter radio emitters than their
black-hole counterparts at the same X-ray luminosity (Fender \&
Kuulkers 2001; Migliari \& Fender 2006; Migliari et al. 2007;
Tudose et al. 2009; Migliari et al. 2010), thus they have been
studied less.

A key point missing in the aforementioned rich phenomenology is a physical
explanation of the jet appearance, disappearance, and
re-appearance as the sources follow a `q'-shaped curve in the
so-called hardness--intensity diagram (HID, first introduced in
the present context by Miyamoto et al.~1995; Fig.~1\footnote{The
figure is publicly available from
http://www.issibern.ch/teams/proaccretion/Documents.html.}). We 
demonstrate that by invoking a simple physical mechanism proposed
more than ten years ago, the so-called Poynting-Robertson cosmic
battery (hereafter PRCB; Contopoulos \& Kazanas 1998), we can
explain in a natural way many of the manifestations of the jets.
Our presentation is semi-quantitative and we postpone detailed
calculations to the future.  Here we simply aim to outline the
general scheme. In Sect. 2 we describe briefly how this mechanism
works, in Sect. 3 we view jet formation and destruction in the
context of the PRCB, and we explain the differences between
neutron-star and black-hole jets, and in Sect. 4 we present our
conclusions and our theoretical predictions.

\begin{figure}
\centering
\includegraphics[angle=0,width=8cm]{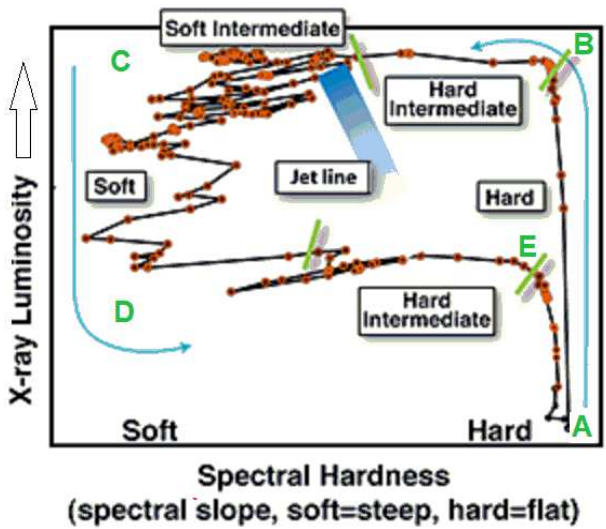}
\caption{ Schematic representation of the `q'-shaped curve in a
hardness-intensity diagram for black-hole X-ray binaries
(see, e.g., Belloni 2010).
} 
\label{Fig1}
\end{figure}

\section{The origin of the magnetic field: A cosmic battery}

The PRCB was originally proposed as a means to account for the
origin of magnetic fields in the accretion flows surrounding
astrophysical compact objects (Contopoulos \& Kazanas 1998; see
also Contopoulos, Kazanas, \& Christodoulou 2006; Christodoulou,
Contopoulos, \& Kazanas 2008). This physical mechanism
has found observational support in observations of the Faraday
rotation-measure gradients across extragalactic jets (Contopoulos
et al. 2009; Christodoulou et al. 2011). The PRCB is based on the
Poynting-Robertson drag effect on the electrons of the innermost
plasma orbiting a black hole or neutron star. We describe here how
the PRCB works in the case of a stellar-mass black hole in an XRB,
and we discuss differences in the case of a neutron star in
\S~3.2.

It is well-known that the accretion disk around a black hole does
not extend all the way to the event horizon, but is truncated at an
inner radius owing to the existence of an innermost stable circular
orbit (hereafter ISCO). As the accretion disk material gradually
spirals in towards the ISCO, it emits blackbody radiation at
increasingly higher temperatures reaching $\sim 10^7$ K at the ISCO.
Within this radius, the material is essentially in free fall and
emits very little radiation.

Electrons in the plasma orbiting the black hole at the inner edge
of the accretion disk scatter photons coming from the opposite
side of the disk inner edge. As they circle the black
hole with a speed comparable to the speed of light, they see the
photons aberrated as if they were 
coming from a slightly head-on direction.
Thus, the electrons experience a force against their direction of
motion\footnote{The aberration of radiation is the essence of the
Poynting-Robertson effect felt by grains in orbit around the Sun.
Contopoulos \& Kazanas (1998) applied the same principle to the
electrons in a plasma orbiting an astrophysical compact object,
assuming an isotropic source of radiation at the center. In what
follows, we relax the assumption of isotropy. A detailed
calculation of the aberration effect in the relativistic
environment of a black hole is in progress.}. The ions feel a much
weaker radiation-drag force than the electrons because the
Thompson cross-section is inversely proportional to the square of
the mass of the scatterer. This difference in the radiation-drag
force leads to an increase in the relative motion between electrons and
ions that is equivalent to an increase in the electric current in the direction
of rotation, hence a stronger poloidal magnetic field. When
one extends the MHD equations to incorporate the equations of
motion of the electron and ion fluids, one sees that the growth of
the axial magnetic field is associated with an azimuthal
electromotive field ${\bf E}_{\rm PR}$ in the direction opposite
that of rotation at the inner edge of the accretion flow
(Contopoulos \& Kazanas~1998). There is little or no azimuthal
electromotive field within the ISCO (where the plasma freely plunges)
or at larger radii in the accretion disk (since the optical depth
increases and the radiation flux decreases rapidly with distance).
The growth of the axial magnetic field around the inner edge of
the accretion flow is driven by the non-zero curl of the
distribution of azimuthal electromotive field ${\bf E}_{\rm PR}$
according to the induction equation
\begin{equation}
\frac{\partial B}{\partial t} = -c\nabla\times {\bf E}_{\rm PR}|_z =
-\frac{c}{r}\frac{\partial}{\partial r}(r E_{\it PR})\ 
\label{induction1}
\end{equation}
where $B$ is the
value of the axial ($z$) magnetic field inside the ISCO, $c$
is the speed of light, and a cylindrical system of coordinates
$(r,\phi, z)$ centered on the central compact object and aligned
with the direction of rotation of the system is used.
We note that the advection and diffusion terms
of eq.~(5) in Contopoulos \& Kazanas (1998) were ignored in
eq.~(1) to avoid complicating the presentation. One can
further simplify eq.~(1) by considering the magnetic vector
potential ${\bf A}$, defined through ${\bf B}=\nabla\times {\bf
A}$, and in particular its $\phi$ component $A_\phi \equiv
\Psi/(2\pi r)$, where $\Psi$ is the total poloidal magnetic flux
accumulated inside radius $r$. Eq.~(1) can then be written as
\begin{equation}
\frac{\partial \Psi}{\partial t} = -2\pi rc E_{\it PR}\ .
\label{induction2}
\end{equation}
Preliminary global axisymmetric ideal MHD numerical simulations
performed by Christodoulou, Contopoulos \& Kazanas~(2008) showed
that this effect does indeed take place around compact
astrophysical objects. The problem with these simulations is that
they need to follow both the short dynamical timescale at the
inner edge of the disk, and the large-scale evolution timescale
for the growth of the magnetic field to dynamically significant
values (see below). To observe the PRCB effect within the
integration time available in our original preliminary simulations
(on the order of a few hundred inner-disk dynamical timescales),
$E_{PR}$ was artificially amplified by several orders of
magnitude. Current supercomputer power allows us, in principle, to
integrate for several million inner-disk dynamical timescales
(Tchekhovskoy, private communication), which will allow us in the
future to perform numerical simulations of the PRCB with realistic
values of $E_{PR}$.

Eq.~(\ref{induction2}) does not take into account the
back-reaction  of the resulting magnetic field on the dynamics of
the plasma in orbit around the compact object. The magnetic field can 
increase in strength to a maximum value dictated by equipartition with the
kinetic energy of the plasma at the inner edge of the accretion
disk, which, for reasons of simplicity, is assumed to lie at
the position of the ISCO of a Schwarzchild (i.e. non-rotating)
black hole (the general case of a spinning black hole in which the
ISCO can lie all the way down to the horizon being currently under
investigation)
\begin{equation}
r = 6GM/c^2\sim 10\left(\frac{M}{M_{\odot}}\right) \mbox{km}.
\label{ISCO}
\end{equation}
We can then estimate the equipartition magnetic field as
\begin{eqnarray}
B_{\it eq} & \sim & (4\pi \rho v_r v_K)^{1/2}\sim
\left(\frac{2\dot{M} v_K }{r h }\right)^{1/2} \nonumber \\
 & \sim & 10^8\ \dot{m}^{1/2}\left(\frac{h}{r}\right)^{-1/2}
  \left(\frac{M}{M_{\odot}}\right)^{-1/2}~\mbox{G}\ ,
\label{Beq}
\end{eqnarray}
where $\rho$ is the density, $v_r$ and $v_K$ are the radial and
the Keplerian velocities, respectively, and $\dot{m}$ is the mass
accretion rate in units of its Eddington value $\dot M_E\equiv
L_E/c^2$, and all quantities have been evaluated at the ISCO.
The above calculation is only one of several possible
ways to estimate $B_{eq}$. Other formulations may consider the
total (gas plus ram) pressure or even the tension of the field
(see footnote 3). These estimates differ from 
those obtained with eq.~(4) by factors of
order `a few'. As described below, our main interest does not
lie in the particular value of the equipartition field, but in the
order of magnitude of the characteristic time that the PRCB takes
to grow the field to near equipartition.

At early times, eq.~(2) applies continuously, and the growth of
the magnetic field is linear with a characteristic timescale
$\tau_{\it eq}$ to reach equipartition equal to
 \begin{eqnarray}
 \tau_{\it eq} & \sim & \frac{r B_{\it eq}}{2cE_{\it PR}}\ .
 \label{tau}
 \end{eqnarray}
To complete our calculation, we need to estimate the PRCB
electromotive field $E_{\it PR}$. As we have said, this calculation
requires a detailed consideration of photon orbits in the
relativistic environment of a black hole. Here we estimate this
term geometrically from the radiation that an electron at the ISCO
sees coming from the `wall' of height $h$ at the opposite side
of the inner edge of the accretion flow. The total luminosity
emitted is
\begin{eqnarray}
 L & \sim & f \sigma T^4 \pi r h \nonumber \\
  & \sim & 10^{36}\ f~ T_7^4 \left(\frac{M}{M_{\odot}}\right)^{2} \left(\frac{h}{r}\right)\
  \mbox{erg}\ \mbox{s}^{-1}\ ,
  \label{L}
  \end{eqnarray}
where $T_7$ is the ISCO equivalent blackbody temperature in units
of $10^7$ K, $\sigma$ is the Stefan-Boltzmann constant, and we include the
geometric factor $f$ of order unity because part
of the radiation coming from the opposite side of the flow 
falls into the black hole and does not reach the other side. The
PRCB electromotive field can now be estimated as
\begin{equation}
E_{\rm PR} \sim g \frac{L \sigma_T}{4\pi r^2 c e}\frac{v_K}{c}\ ,
\label{EPR}
\end{equation}
where $\sigma_T$ is the electron Thomson cross-section, $v_K$ is
the Keplerian velocity at the ISCO, $v_K/c$ is the
Poynting-Robertson aberration effect caused by photons `hitting'
the moving electrons on the other side at about $90^\circ$, and
$g$ is a geometric factor of order unity that accurately
accounts for light bending in the immediate environment of the
central black hole. Substituting these expessions into eq.~(5),
we obtain
\begin{equation}
\tau_{\it eq} \sim 4\times 10^5 (fg)^{-1} T_7^{-4}\dot{m}^{1/2}
\left(\frac{h}{r}\right)^{-3/2}
\left(\frac{M}{M_{\odot}}\right)^{1/2}\ \mbox{s} . \label{tau2}
\end{equation}
We recall that Contopoulos \&
Kazanas~(1998) obtained a simpler estimate of the above timescale
under the simplifying assumption that the source luminosity
originates from the center, namely $\tau_{eq}\sim 4\times 10^{-5}\
\dot{m}^{-1/2}(M/M_{\odot})^{3/2}$ ~yr. Our present, more elaborate
calculation, elucidates the dependence of the PRCB timescale on the
physical state of the accretion disk (thick versus thin), which seems
to be of paramount importance in the phenomenon of jet formation
and destruction in XRBs (see Sect. 3).

To reach a current strong enough to produce the magnetic
field on the order of $10^6$ G ($\sim 10$\% of the equipartition
value) required to explain the radio emission in a $10M_{\odot}$
source such as Cyg X-1 (Giannios 2005), the battery must operate for
about 12 hours, if the inner part of the accretion flow is
described by an ADAF with $h/r\sim 1$ and $\dot m \sim 10^{-1}$.
This timescale is short compared to the system evolution timescale
up to the first crossing of the jet line 
(see Fig. 1) or after the second
crossing of the jet line at lower X-ray luminosity (as an example,
we refer to the observed timescales for the 300-day outburst of J1752-223
in Stiele et al. 2011).

However, when the system moves from the hard to the
soft state, the accretion flow changes from a thick ADAF to a
Shakura-Sunyaev thin disk with $h/r\sim 10^{-1}$. In this case,
the timescale for the growth of a $10^6$ G magnetic field ($\sim
3$\% of the equipartition value) becomes on the order of five days
(assuming that $\dot{m}$ is still $\sim 10^{-1}$), which is longer than
the characteristic evolution timescale of the black-hole XRB in
the soft state (see the evolution past region C in Fig. 1, Sect. 3.1
below, and Fig. 4 of Stiele et al. 2011).

\section{Jet formation and destruction}

A vast amount of work done in three-dimensional (3D) 
magnetohydrodynamic (MHD) simulations of astrophysical
accreting systems (e.g. Machida, Inutsuka \& Matsumoto~2006;
Hawley~2009; Romanova et al.~2009; Mignone et al.~2010 and
references therein) have apparently shown that jet formation and
acceleration is due to one of two types of physical mechanisms:
plasma gun/magnetic towers, as proposed by
Contopoulos (1995)/Lynden-Bell (1996), or centrifugal driving, as
proposed by Blandford \& Payne (1982). What is most important,
however, is that these simulations demostrate that at the origin of jet
formation and acceleration lies a large-scale magnetic field that
threads the central `driving engine' (the central compact object
and the surrounding innermost accretion disk). Once this field
is in place, jet formation can proceed through any one of the
above mechanisms. In contrast, without an established 
large-scale magnetic field, jet formation is impossible.

Most early simulations of MHD jet formation did not investigate
the origin of the required large-scale magnetic field
(it was simply assumed as an initial/boundary condition).
Were the disk material to be infinitely conducting, accretion would
naturally advect and amplify any weak magnetic field present on
large scales. Magnetorotational instability
(MRI) simulations, however, suggest that astrophysical
accretion disks develop turbulent magnetic diffusivities
comparable to their turbulent viscosities, indicating that weak
large-scale magnetic fields rapidly diffuse outward (e.g. Lubow,
Papaloizou, \& Pringle~1994). Numerical simulations have
shown that, under certain physical conditions,
advection succeeds in establishing a strong magnetic field around the
central object (e.g. Igumenshchev~2008; Lovelace, Rothstein, \&
Bisnovatyi-Kogan~2009; Tchekhovskoy, Narayan, \& McKinney~2011).
We are currently investigating these conditions.

In the present paper, we consider an independent
physical scenario and associate the origin of the magnetic field
with the PRCB. In what follows, we assume that, whenever the
PRCB operates efficiently, a large-scale magnetic field is
generated in the immediate vicinity of the central compact object,
which, according to previous 3D MHD simulations, leads to the
formation of jets. In the opposite limit, when the PRCB operates
inefficiently, the resulting magnetic field is weaker, as is
the jet power and emission.
Therefore, under the assumption that the PRCB is in
operation in XRBs, we examine whether a consistent picture can be
drawn that explains all the manifestations of the jets in these
sources.

\subsection{Jets in black-hole X-ray binaries}

In the HID, a steady jet exists even when the sources are in the
so-called quiescent state, where the spectrum is dominated by 
hard X-rays, and typically $L_X \sim 10^{-8} - 10^{-6} L_E$ (Kong
et al. 2002; Hameury et al. 2003; Remillard \& McClintock 2006;
Gallo et al. 2006). To be more accurate, the `q'-shaped curve
shown in Fig. 1 may turn to the left at its bottom right part,
i.e. the hardness ratio may be somewhat lower than at
higher intensities (Corbel, K\"ording, \& Kaaret 2008; Cabanac et
al. 2009). Since the sources spend most of their time in this
state, the PRCB mechanism has enough time to establish a sufficiently strong
magnetic field to permit the ejection of a jet. This may not be the case
though for all black-hole XRBs in the quiescent state. For
example, for GRO J1655-40 and XTE J1550-564, upper limits on the radio
emission in the quiescent state were placed (Calvelo et al. 2010).
In the source V404 Cyg, the radio emission is, in contrast, stronger
than expected in the quiescent state (Gallo,
Fender, \& Hynes 2005). As a rule however, black-hole XRBs exhibit
a radio jet in the quiescent state and the PRCB can account for
it.

In a typical outburst, as the X-ray luminosity increases, the
hardness ratio remains approximately constant, and in the HID the
sources trace the vertical line of the `q'-shaped curve (region
marked as hard state in Fig. 1).  A steady, compact, partially
self-absorbed jet (first imaged in Cyg X-1 by Stirling et al.
2001) is present in this part of the `q'-shaped curve (region B in
Fig. 1) and its radio luminosity $L_R$ also increases in
correlation with the X-ray luminosity $L_X$ as $L_R \propto
L_X^{0.71 \pm 0.1}$ (Corbel et al. 2000) or more accurately as
$L_R \propto L_X^{0.6} M^{0.8}$, where $M$ is the mass of the
black hole (Merloni, Heinz, \& di Matteo 2003; Falcke, K\"ording,
\& Markoff 2004). The jet is mildly relativistic with a bulk
Lorentz factor $\gamma < 2$ (Fender, Belloni, \& Gallo 2004).
Since the hard X-rays dominate the luminosity, this state is
called hard. In addition, because the accretion disk is radiatively
inefficient, it is generally believed to have an advection
dominated accretion flow (ADAF) (Narayan \& Yi 1994; 1995) in its
inner part (Gierlinski, Done, \& Page 2008; Cabanac et al. 2009).
The ADAF is geometrically thick, optically thin, and quite hot.
The ADAF and the jet are the most likely origins
of the hard X-ray spectrum (Giannios 2005; Markoff,
Nowak, \& Wilms 2005). As the X-ray luminosity of the sources
increases, the electric current produced by the PRCB in the inner
ADAF and the magnetic field in the steady jet increase. Thus,
the radio emission also increases (e.g., Giannios 2005). It is then
unsurprising that the radio luminosity and the X-ray luminosity
are correlated.

As the sources reach the upper right part of the `q'-shaped curve
(region B in Fig. 1), the X-ray luminosity is at its maximum and
the steady jet is at full strength.  This is because the inner
part of the accretion disk is occupied by the ADAF and the
geometrically thin disk extends from some radius outwards (Cabanac
et al. 2009). The PRCB works at its highest efficiency.

The hard state and the steady jet persist as the sources trace the
upper right part of the `q'-shaped curve (region to the left of B
in Fig. 1, marked as the hard intermediate state). The spectrum
softens somewhat, but little else changes.

At more or less constant intensity, the hardness ratio then
decreases further during the outburst and approaches the so-called
jet line, which is the line between the existence and the non-existence of
a compact jet.  As the source approaches the jet line, the radio
emission peaks and the jet properties change drastically (Fender,
Homan, \& Belloni 2009). The emission is no longer steady but
eruptive, emitting discrete blobs, and its bulk Lorentz factor
increases to $\gamma >2$.  The high radio flares occur in most
sources for extended periods of time (GX 339-4 is a notable
exception) and are also associated with strong X-ray flaring 
(Brocksopp et al. 2002; Fender, Homan, \& Belloni 2009). The
physical picture we have in mind is the following: as the outburst
evolves, $\dot m$ apparently increases, the geometrically thin
disk extends further in, less room is left for the ADAF, and the
ADAF gradually disappears. As the sources reach the jet
line, the geometrically thin disk extends all the way to the ISCO.
Regardless of the magnetic field that existed inside the ISCO of the
geometrically thick ADAF will now have trouble being held there by
the thin disk. We propose that the thin disk becomes unstable to
non-axisymmetric magnetic `Rayleigh-Taylor-type' instability
modes, and the accumulated magnetic field escapes to the outer
disk in the form of magnetized `strands'\footnote{While in the
ADAF stage, magnetic pressure is balanced by the infall ram
pressure (eq.~4). After plasma cooling and the collapse of the
disk to $h/r \simless 10^{-1}$, the magnetic stresses at the inner
edge $B_z dB_r/dz \sim B_z B_{r,surface}/h \sim B_z^2/h$ become
too strong for the thin disk to contain the field and 
instabilities are expected to set in.}. Such an instability may
explain naturally the flaring nature of the jets as the sources
approach the jet line during the hard-to-soft transition.

To the left of the jet line (regions C and D in Fig. 1), the
sources are in the so-called soft state, the accretion disk is
widely believed to be of the Shakura-Sunyaev (1973) type, i.e.,
geometrically thin and optically thick, and extends all the way to
the ISCO.  In this state, radio emission is in nearly all cases
either undetectable or optically thin, consistent with the
disappearance of the compact, partially optically thick, steady
jet that was present in the hard state (Fender, Homan, \& Belloni
2009).  The black-hole XRB H1743-322 provides the tightest constraints
to date for radio quenching in the soft state (a jet quenching
factor of $\sim 700$; Coriat et al. 2011). The PRCB works in this
state also, but since the thickness of the disk $h$ is much
smaller than the radius $r$ of the ISCO, the efficiency of the
battery is significantly lower.  For example, for $h/r \approx
10^{-1}$, the timescale for the establishment of an equipartition
magnetic field at the inner part of the thin disk (eq. 8 above) is
on the order of five months at the upper left part of the `q'-shaped
curve (region C in Fig. 1) and becomes even longer as the
intensity of the source decreases (transition from region C to
region D in Fig. 1)\footnote{From eqs.~(5) \& (7), the timescale
for field growth is inversely proportional to the XRB X-ray
luminosity.}. This timescale is substantially longer than the
characteristic time that the sources spend in the various parts
in-between regions C and D in Fig. 1.

Some sources, including the archetypal black-hole XRB GX 339-4,
cross the jet line only once (Fender, Homan, \& Belloni 2009).
While in the soft state, they may trace small closed loops (see,
e.g., Fig. 7 of Fender, Belloni, \& Gallo 2004). No jet has ever
been detected when the sources trace these loops.  However,
when a source does decide
to go back and forth and thus to cross the jet line
leisurely several times, then
a) a steady jet is created as the hardness ratio
increases (soft-to-hard state transition -- the inner accretion
disk gradually thickens, the magnetic field is built at an
accelerated rate, and the disk becomes stable to non-axisymmetric
Raleigh-Taylor modes); b) a flaring jet appears as the sources
return to the jet line (hard-to-soft state transition -- the inner
disk becomes thinner, and the accumulated field makes the thinner
disk unstable); and c) just as in the first crossing of the jet
line, the jet disappears in the soft state. This is exactly what
is observed in GRS 1915+105 (Rushton et al. 2010; see also
Brocksopp et al. 2002 for XTE J1859+226). This source is
known to be in the soft state for long periods of time, although
it makes repeated excursions to the hard state.

Exactly the opposite seems to occur in Cyg X-1 (Rushton et
al. 2011). This source is known to be in the hard state for long
periods of time (Di Salvo et al. 2001), although it makes
repeated attempts (`failed' state transitions; Pottschmidt et
al. 2003) to reach the soft state, where it did exist at least
once (Fender et al. 2006).

To complete the `q'-shaped curve, the sources in the decay phase
of the outburst increase their hardness ratio at about constant
intensity and cross the jet line from the left (region marked 
hard intermediate state, to the left of E in Fig. 1).  The intensity at
this crossing (soft-to-hard state transition) is significantly
lower than in the previous crossing (hard-to-soft transition). The
accretion flow again becomes ADAF, initially at the outer part and
then at a larger extent, and the PRCB now operates very
efficiently. As a result, a jet gradually forms (Brocksopp et al.
2005; Fender, Homan, \& Belloni 2009), and by the time that the
sources reach region E of Fig. 1, a steady jet is established and
persists as the intensity declines and the sources transit to the
quiescent state (region A of Fig. 1).

\subsection{Jets in neutron-star X-ray binaries}

Low-magnetic-field neutron-star XRBs are divided into two distinct
classes, based on their X-ray spectral and timing properties. They
are named atoll and Z sources (Hasinger \& van der Klis 1989)
because of the shape that they trace in a color-color diagram.
Both classes exhibit jets, but because their radio emission is
weaker than that in black-hole XRBs, they have received less
attention (for comprehensive reviews of the properties of jets
from neutron-star XRBs, see Migliari \& Fender~2006 and K\"ording,
Fender, \& Migliari~2006). Moreover, neutron stars are more
complex than black holes in that they possess a solid surface and
a stellar magnetic field. For the above reasons, our discussion of
jets in neutron-star XRBs here is more tentative and qualitative
than the discussion of Sect. 3.1.

In the HID, neutron-star XRBs trace a `q'-shaped curve, similar to
that of black-hole XRBs (Maitra \& Bailyn 2004).  Nevertheless, there are
two major differences between the jets in these two types of
binaries:
\begin{enumerate}
\item The radio luminosities of neutron-star jets are typically
$\sim 30$ times lower than those of black-hole jets at
comparable X-ray luminosities (Fender \& Kuulkers 2001; Migliari
et al. 2003). \item Radio emission has been detected in the soft
state of some neutron-star XRBs (Migliari et al. 2004), as opposed
to their non-detection in black-hole sources, when they are in the
same state.
\end{enumerate}
Both of these differences can be explained physically using the
PRCB along with the fact that neutron stars have a solid surface.
Unlike in black-hole XRBs, where the accretion disk extends inward
to the ISCO, in neutron-star XRBs with weak magnetic fields 
the accretion disks extend all the way to the surfaces of the
stars because in this case, the ISCO practically coincides with
the stellar surface.

One of the well-studied neutron-star binaries in both radio and
X-rays is the atoll source Aql X-1.  The evolution of the radio
and the X-ray emission of this source throughout an entire
outburst was reported  by Miller-Jones et al.
(2010). In the HID, Aql X-1 traces a `q'-shaped curve, similar to
that of black-hole XRBs. In the hard state, the source exhibits a
compact jet, similar to those of black-hole XRBs, but with weaker
radio emission.  The inner part of the accretion disk is of the
ADAF form, as in black-hole XRBs, and thus the PRCB operates in
some capacity.  The difference is that now the solid surface of
the neutron star blocks the radiation from the opposite side of
the inner accretion disk.
Thus, the PRCB now operates at a reduced efficiency.
In addition, the radio emission is weaker for the following
important reason. The
X-ray luminosity scales as $L_X(NS)\propto \dot{m} M$ for all
neutron-star states and as $L_X(BH)\propto \dot{m}^2 M$ for black
holes accreting in the ADAF state (Narayan, Garcia, \& McClintock
1997; Abramowicz \& Fragile 2011), whereas the radio luminosity scales as
$L_R\propto \dot{m} M$ in both cases; hence for accreting
sources of comparable X-ray luminosities, the ratio of their jet (radio)
powers will be $L_R(BH)/L_R(NS)\simeq \dot m^{-1}$ .

Although neutron-star XRBs can produce ultra-relativistic outflows
in some cases (Fender et al. 2004), their transition from
the hard to the soft state generally occurs in a much smoother way
than in black-hole XRBs (Miller-Jones et al. 2010).  This can also be
understood physically by the PRCB continuing to
operate, albeit at a lower efficiency, when the neutron-star
sources transit from the hard to the soft state.

In the soft state, the jet of Aql X-1 is quenched.  This is expected
because in the soft state the inner accretion disk is thinner
(Shakura-Sunyaev type),
and this part of the disk as well as
the region where accreting matter impinges onto the neutron-star surface
(Inogamov \& Sunyaev 2010) are both optically thick.  This implies that 
there is little difference in the azimuthal velocities between
the points of emission and scattering of the radiation.
Thus, the PRCB is quite inefficient.

Unlike black-hole XRBs, radio emission has been detected in some
cases in the soft state of neutron-star XRBs (Migliari et
al. 2004). These cases can be understood as follows.  In the area
where the accreted material impinges onto the neutron-star
surface, a `spreading layer' likely forms with scale height much
larger than the actual thickness of the accretion disk (Inogamov
\& Sunyaev 2010). Radiation from this `spreading layer' then hits
the accretion disk from above and below, thus fueling the PRCB.
The battery operates in this case too, though at a
lower efficiency than for when the inner part of
the accretion disk is of the ADAF type.  (Here we have made the tacit
assumption that there is a non-negligible difference between the
spin frequency of the neutron star and the Keplerian frequency of the disk.)

\subsection{Additional remarks}

X-ray pulsars do not exhibit compact jets despite their strong magnetic
fields (a few times $10^{12}$ G).  In the context of the PRCB, this
can be understood as follows. The accretion disks extend inwards
to the Alfv\'en radius, which is on the order of $10^8$ cm
(Ghosh \& Lamb 1978). At this radius, and because the accretion
disks are geometrically thin, the photon flux at the inner edges of
the disks is very low and the PRCB is thus very inefficient.
In addition, the accretion flow distorts the overall dipolar geometry of the
magnetic flux generated by the PRCB since, in an X-ray pulsar, accretion
does not proceed radially toward the center, but follows magnetic field
lines toward the magnetic poles.

However,
it is unclear what the magnetic-field strength is in atoll and
Z sources.  It is generally thought to be low because
the discovery (Wijnands \& van der Klis 1998) of coherent 401 Hz X-ray
pulsations from the accretion-powered pulsar SAX J1808.4-3658
allowed the determination (Psaltis \& Chakrabarty 1999) of a surface
magnetic field of $10^8 - 10^9$ G in this source.  However, it is unknown
whether other atoll and Z sources have surface magnetic fields of this 
order of magnitude; their magnetic fields could very well be weaker.

If the surface magnetic field of neutron stars is weaker than
the equipartition value
given by eq.~(4) above, then the PRCB operates effectively.  On the other hand,
if the surface magnetic field is stronger than the equipartition value,
this strong magnetic field determines where the inner disk
will be truncated and whether a jet will be produced.
Judging from the observed X-ray pulsars, it seems that a significant
surface magnetic field inhibits jet formation.

Along the same line of reasoning, it is unsurprising that SS
Cyg, a non-magnetic white-dwarf XRB, exhibits a `q'-shaped curve
in the HID and jet emission characteristics similar to the atoll
source Aql X-1 and the black-hole XRB GX 339-4 (K\"ording et
al. 2008).  Thus, it appears very likely that atoll and Z
sources, as a class, have very weak magnetic fields.

\section{Conclusions}

Not only is the PRCB able to explain the poloidal magnetic fields required
for the ejection of jets in white-dwarf, neutron-star, and
black-hole XRBs, but it also provides a physical framework through
which we can understand many of the jet manifestations in these
sources in relation to their spectral states. Equally importantly, the PRCB
provides naturally timescales for these transitions that are much longer
than the dynamical ones and in general agreement with those
observed. Our understanding is semi-quantitative at this point,
but detailed calculations are underway and will be reported
elsewhere.

In view of our proposed picture, we make the following predictions about
black-hole XRBs:

\begin{enumerate}
\item The transition from the hard to the soft X-ray
state (first jet-line crossing in the HID) will {\it always} be associated
with eruptive
jet emission, flaring, and the disappearance of the compact jet past the
jet line.

\item The transition from the soft to the hard X-ray state (last
jet-line crossing in the HID) will {\it always} be associated with
the formation of a steady jet within $1 - 5$ days of the crossing.
No eruptive emission will {\it ever} be seen in such a one-way
crossing.

\end{enumerate}

In the case of neutron-star XRBs, our predictions are the following:

\begin{enumerate}
\item No compact jet will be formed in normal X-ray pulsars (magnetic
field $\simmore 10^{12}$~G).

\item No jet will be formed in the soft state of atoll and Z sources
if the spin frequency of the
neutron star is comparable to the Keplerian frequency of the inner
part of the accretion disk.

\item In atoll and Z sources, the larger the difference between the
neutron-star spin frequency and the
Keplerian frequency of the inner part of the accretion disk, the stronger
the radio jet in the soft state\footnote{This prediction was
recently confirmed by Migliari, Miller-Jones, \& Russel (2011).}.
\end{enumerate}

\begin{acknowledgements}
This research has been supported in part by EU Marie Curie project
no. 39965, and EU REGPOT project number 206469. NDK thanks Tomaso
Belloni for providing hardness-intensity data of GX 339-4 during
various outbursts and Michiel van der Klis and Dimitrios Psaltis
for useful discussions concerning millisecond pulsars. IC thanks
Maxim Lyutikov, Ramesh Narayan, Roman Shcherbakov, and Sasha
Tchekhovskoy for discussions concerning accretion disks and the
PRCB.

\end{acknowledgements}

\end{document}